\documentclass[aps,prd,reprint,showpage,showpacs,nobibnotes,superscriptaddress,twocolumn,amssymb,amsmath,nofootinbib,floatfix]{revtex4-1}

\usepackage{graphicx,amsfonts}
\usepackage{epsfig}
\usepackage{dcolumn}
\usepackage{bm}
\usepackage{color}
\usepackage{soul}
\usepackage{siunitx}
\usepackage{subcaption}
\begin{document}

\title{Trimuon production at the LHC}

\author{Mario W. Barela}
\email{mario.barela@unesp.br}
\address{
Instituto de F\'\i sica Te\'orica, Universidade Estadual Paulista, \\
R. Dr. Bento Teobaldo Ferraz 271, Barra Funda\\ S\~ao Paulo - SP, 01140-070,
Brazil}

\author{V. Pleitez}
\email{v.pleitez@unesp.br}
\address{
Instituto de F\'\i sica Te\'orica, Universidade Estadual Paulista, \\
R. Dr. Bento Teobaldo Ferraz 271, Barra Funda\\ S\~ao Paulo - SP, 01140-070,
Brazil}

\begin{abstract}
No process without a standard model (SM) background has been observed so far. As a tool in the study
of a class of models containing doubly charged vector bileptons, we propose one such process that,
violating lepton flavor number conservation, has no contribution from the SM: $pp\to \mu^+\mu^+\mu^- e^-$. By
carefully isolating the parameters to keep them free, we are able to acquire a notion of how a possible
PMNS-like matrix present in the relevant charged current parametrization could affect the observables. We
find that an interesting section of the space of parameters can be explored at the current LHC in the
specified conditions.
\end{abstract}

\maketitle

\section{Introduction}
\label{sec:intro}

The Large Hadron Collider (LHC) has reached an
excited stage of a large amount of data collection. Even
though conclusive hints for new physics have been evasive,
current data and the data that are expected to be collected at
the high-luminosity (HL) stage open the possibility of
turning the LHC into a precision machine.

In this paper, we explore the prospects for the LHC to
discover doubly charged vector bileptons with the current
luminosity. Most of the processes considered to find new
physics have a SM background. This is not the case for the
process $pp\to \mu^+\mu^+e^-e^-(e^+e^+\mu^- \mu^-)$~\cite{Meirose:2011cs}, or the trimuon events $pp\to \mu^+\mu^+\mu^-e^-$, both of which violate the con-
servation of lepton flavor number, which is conserved to all
orders in perturbation theory within the SM. Hence, having
no background, this sort of process may be the smoking
gun of new physics and, specifically, of the discovery of
doubly charged particles, which are, in many models, the
biggest candidates to trigger these processes.

Doubly charged particles appear in multiple scenarios
beyond the standard model (BSM) with extended gauge
groups; they may be scalars, fermions, or vectors: see
Ref.~\cite{Alloul:2013raa} and references therein. 
Among the possibilities, the more interesting one is the case of doubly charged \textit{vector} bosons, because (i) their couplings with leptons have almost the same intensity as the $W^\pm$ of the SM, and (ii) this kind of particle is a very rare feature in models of new physics. They occur, for example, in the minimal 3-3-1 model (m331 for short)~\cite{Pisano:1991ee,Frampton:1992wt,Foot:1992rh}, and in $SU(15)$ grand unification~\cite{Frampton:1989fu,Frampton:1991mt}. 

If these sorts of particles do exist, then resonances in like-sign leptons' invariant mass could be observed at the LHC~\cite{Coriano:2018khp} in the (sub)process $U^{++} \to  \ell^+\ell^+$.  Interesting cases include when $pp\to e^+e^+\mu^-\mu^-$~\cite{Meirose:2011cs} and when $pp\to \mu^+\mu^+\mu^-\mu^-$~\cite{Dion:1998pw,Nepomuceno:2016jyr}.

This paper is organized as follows: In Sec.~\ref{sec:int}, we write down the interactions relevant for our analysis. In Sec.~\ref{sec:lhc}, we describe the method and present our results.  Our conclusions are presented in Sec.~\ref{sec:con}.

\section{Interactions}
\label{sec:int}

 The lepton-lepton-bilepton interaction that is relevant to the present analysis is the following:
\begin{equation}
\mathcal{L}_{\ell \ell}=
-i\frac{g}{4\sqrt2}\bar{\ell}^c\gamma^\mu(A-\gamma_5B)\ell U^{--}_\mu+ \text{H.c.}.
\label{llu}
\end{equation}

For general $A$ and $B$ matrices, this is a model independent parametrization for vector and axial interactions.
We will focus here on a large class of models in which the charged lepton mass matrix is diagonalized by a biunitary transformation given by $\hat{M}^{\ell} =V^{\ell \dagger}_L M^\ell V^\ell_R$, defining $\ell^\prime_{L,R}=V^\ell_{L,R}\ell_{L,R}$ and $\hat{M}^\ell=\textrm{diag}(m_e,m_\mu,m_\tau)$, where the primed fermions are symmetry eigenstates and the unprimed ones are mass eigenstates. In this scenario, $A$ is an antisymmetric matrix given by $A=V_U-V^T_U$, while $B$ is symmetric and obeys $B=V_U+V^T_U$, where $V_U=(V^\ell_R)^TV^\ell_L$.

In a model independent way, the significant production mechanism of the bileptons is through Drell-Yan-like processes. We will need the bilepton-bilepton-$Z$ interaction, generally given by
\begin{eqnarray}
\mathcal{L}_{UUZ}&=& i\frac{g}{2}f(g,v)\{ U^{++}_\mu [U^{--}_\alpha (\partial_\mu Z_\alpha )-Z_\alpha (\partial_\mu U^{--}_\alpha )] 
\nonumber \\&+&  U^{--}_\nu [Z_\alpha (\partial_\nu U^{++}_\alpha )-
U^{++}_\alpha (\partial_\nu Z_\alpha )]  
\nonumber \\&+&  
Z_\alpha [ U^{++}_\mu (\partial_\alpha U^{--}_\mu )-U^{--}_\mu (\partial_\alpha U^{++}_\mu )]\}; 
\label{uuz}
\end{eqnarray}
where $f(g,v)$ is a dimensionless function of the gauge
coupling constants and of the vacuum expectation values of
the model. Distinctive possible values for $f(g,v)$ include (i) $f(g,v)=2c_W$, which is the corresponding value of the SM $W^+W^-Z$ vertex, and (ii) $f(g,v)=-(1-4s^2_W)/c_W$, which is the vertex of the m331 model in a possible SM limit~\cite{Dias:2006ns} if we use $g^2_X/g^2=s^2_W/(1-4s^2_W)$, where $g_X,g$ are the gauge coupling constants of $U(1)_X$ and $SU(3)_L$, respectively, and $s_W, c_W$ are the sine and cosine of the the weak angle. Below we will consider only the latter case. Notice that in the m331 model, the vector bilepton if $Z$-phobic. 

The last required lagrangian is that of the bilepton-bilepton-photon interaction, which is:
\begin{eqnarray}
\mathcal{L}_{UUA}&=&i2Q_e(g)\{ U^{++}_\mu[U^{--}_\alpha (\partial_\mu A_\alpha)-A_\alpha(\partial_\mu U^{--}_\alpha)]  \nonumber \\&+&U^{--}_\nu[A_\alpha(\partial_\nu U^{++}_\alpha)-U^{++}_\alpha(\partial_\nu A_\alpha)]\nonumber \\&+&
A_\alpha [U^{++}_\mu(\partial_\mu U^{--}_\mu)-U^{--}_\mu(\partial_\alpha U^{++})]\};
\label{uua}
\end{eqnarray}
here $Q_e(g)$ is the expression for the fundamental charge within the considered model, which, in general, is a function of the coupling constants. In our calculations, we will set $Q_e(g)=gs_W$, which is the corresponding electrical charge of the SM and also of the m331 when using, again, $g^2_X/g^2=s^2_W/(1-4s^2_W)$.

\section{LHC phenomenology}
\label{sec:lhc}

As mentioned before, we will focus on the phenom-
enology of the vector bileptons at the LHC energy and
luminosity. We note that in all the previous analysis, only a
diagonal version of Eq.~(\ref{llu}) has been considered, and
consequently, the trimuon final-state case was not specifi-
cally studied either. This may be too restrictive an impo-
sition, since, taking the m331 model as an illustration, it is
not possible to assume that the charged lepton mass matrix
is in the diagonal basis. This is because to generate the
correct mass of these particles, it is necessary to have two
contributions arising from different scalar multiplets: a
triplet $\eta$ and a sextet  $S$. The mass matrix coming from the
Yukawa interactions between the leptons and the triplet is
antisymmetric, while that from the sextet is symmetric. If
we choose only the symmetric matrix (forgetting the
sextet), the neutrino mass matrix becomes proportional
to the charged lepton one, so that they are diagonalized by
the same transformations, which, in turn, causes the
resulting PMNS matrix to be unity.

In turn, this theoretically (probably) inescapable mixing causes the study of this processes to be very difficult as a consequence of the number of free parameters. For this
reason, we perform a study of the trimuon end state with
more general nondiagonal mixing matrices, considering
only the contribution of the bilepton $U^{\pm \pm}$ together with the
SM particles. Of course, we stress that in this case the
unitarity of the model is not manifest, but we already know
possible ultraviolet completions, say those in Refs.~\cite{Pisano:1991ee,Frampton:1992wt} or \cite{Frampton:1989fu}. Eventually, directed studies of specific models should take all contributions into account.

\subsection{The method}

In order to obtain exclusion contours in the two-dimensional parameter space $M_U \times (V_U)_{e\mu}$, where $V_U$ is the unitary matrix introduced below Eq.~(\ref{llu}), we study the process $pp\to e^- \mu^- \mu^+ \mu^+$ (and its charge mirrored end-
state conjugate), which, obviously, has no SM background,
and hence may be easily distinguished if it does happen at
all in the current experimental reach. However, further
simplifications are needed if we are to be able to perform
this study without arbitrarily fixing unknown parameters.  

\begin{figure}[b]
	\includegraphics[width=0.68\linewidth]{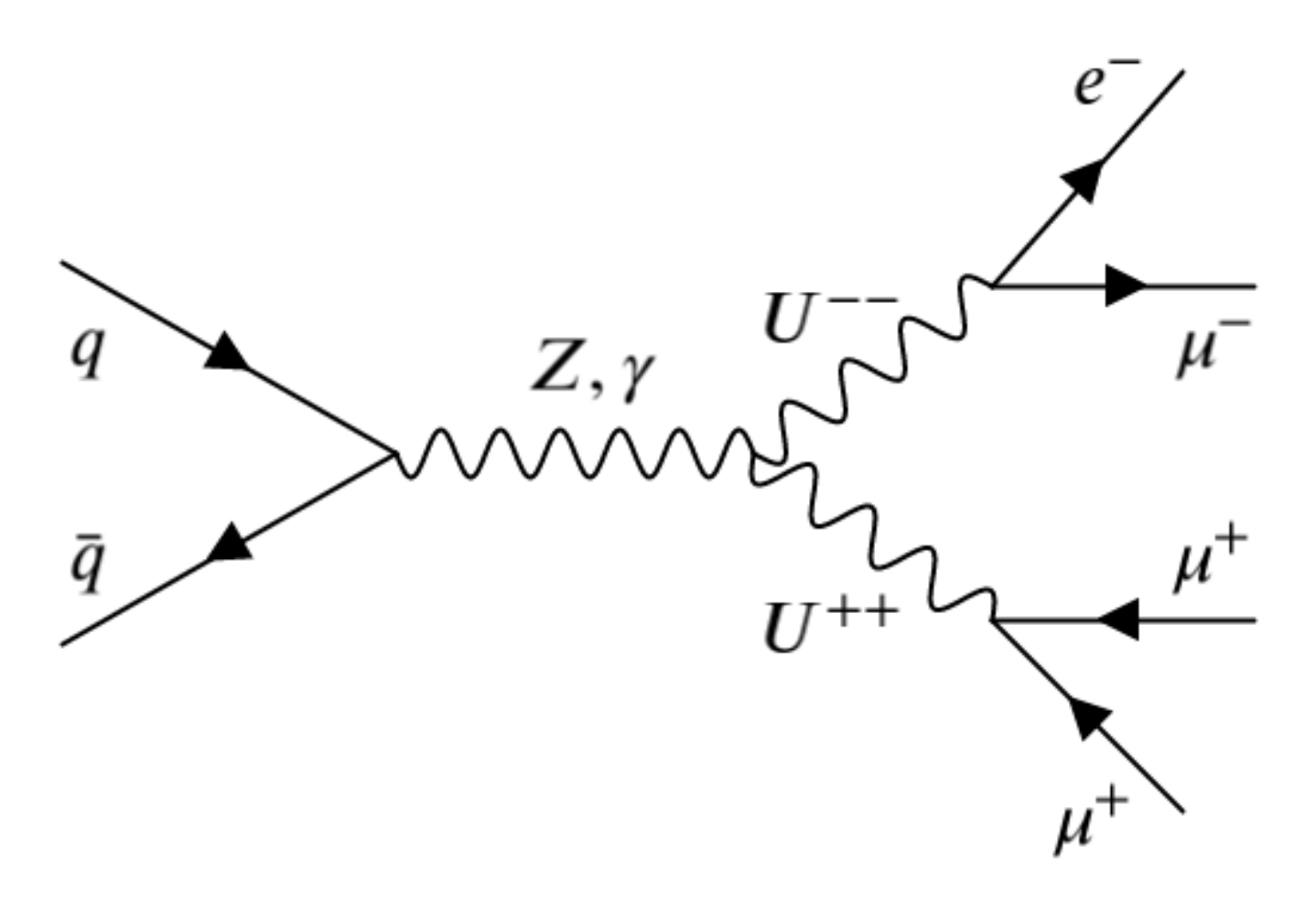}  
	\includegraphics[width=0.8\linewidth]{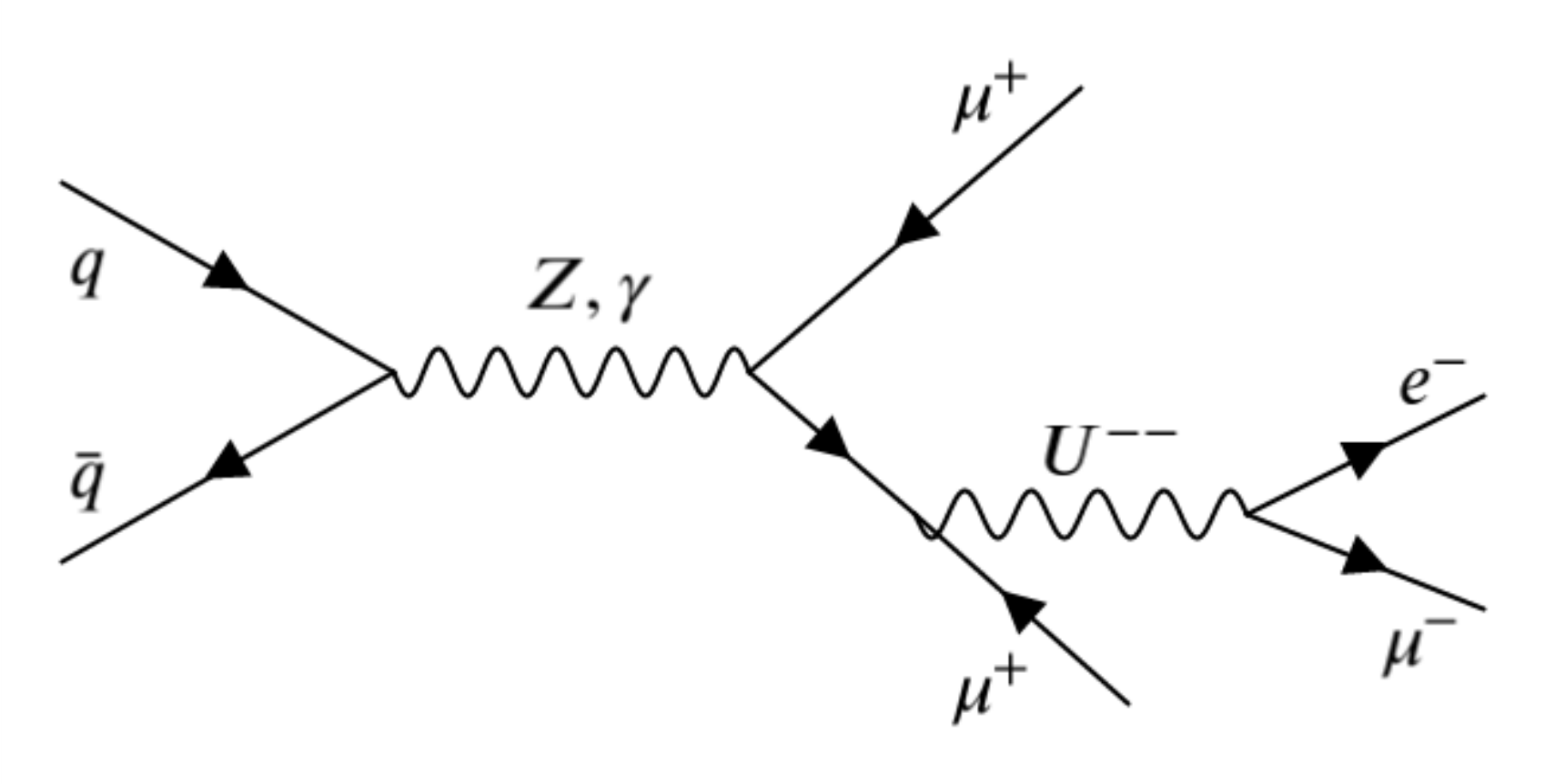} 
	\caption{Representatives of the considered diagrams.}
	\label{fig2}
\end{figure}

We start by recalling that we are considering only contributions containing the $U^{\pm \pm}$ and the SM particles. Diagrams are shown in Fig.~\ref{fig2}. 

At this point, we would be left with five free parameters:
the boson’s mass and the three angles and one phase that
parametrize a $3 \times 3$ unitary matrix -- all the matrix elements
influence the results, if not explicitly in the vertex, through
the particle’s width. The next obvious choice of simplifi-
cation is to make the matrix real, eliminating in this way the
phase. At last, we decide to study the case of a symmetric $V_U$, ending up with three parameters to deal with: the mass
and the two real numbers that are the degrees of freedom
(d.o.f.) of a $3 \times 3$ symmetric orthogonal matrix (actually,
there is still one more discrete d.o.f. that labels one of four
different solutions of the orthogonality conditions for the four other matrix elements in terms of two specified ones,
but this is not an issue). We stress that these assumptions,
although not more strong than necessary for any similar
analysis in the present phenomenological context, are not
the most general case or a prediction of a specific model. 

A consequence of imposing a symmetry condition on $V_U$ is that there are no vectorial interactions between the bilepton and the leptons, since, again, $A=V_U-V_U^T$.

Our goal is to learn what is the behaviour of the signal upon the variation of the parameters $m_{U^{\pm \pm}}$ and $(V_U)_{e\mu}$. To do so, we chose the third free parameter from the analysis above to be $(V_U)_{ee}$, which we fix at four different benchmark values: $(V_U)_{ee}=\{0.001,0.01,0.1,0.9\}$. We then perform one bidimensional scan for each value of $(V_U)_{ee}$, for values within $\SI{100}{GeV} \leq m_{U^{\pm \pm}} \leq \SI{1200}{GeV}$ in steps of $\SI{50}{GeV}$ and $0.001 \leq (V_U)_{e\mu} \leq 0.9$ -- except for $(V_U)_{ee}=0.9$, when $(V_U)_{e\mu}$ goes up to $\sim 0.43$ -- through 12
strategically chosen points, with every other matrix element
being in each point determined as a solution of the
orthogonality constraints. Using the Monte Carlo generator \texttt{MadGraph}, we generated 10\,000 events for each of the 1035 points in the parameter space, with the following cuts on transverse momentum, rapidity, and opening angle between leptons:

\begin{equation}
\SI{1500}{GeV}>p_{T_\ell}>\SI{30}{GeV}, \;\; |\eta_\ell|<2.5, \;\; \Delta R_{\ell \ell}>0.4
\end{equation}
at a center-of-mass energy $\sqrt{s}=\SI{13}{TeV}$ and an integrated luminosity of $\mathcal{L}=\SI{140}{fb^{-1}}$. The resulting total cross section of each point is multiplied by 2, to accommodate the charge-reversed end state $p p > e^+ \mu^+ \mu^- \mu^-$, which has identical numerical results in our case of $(V_U)_{ij}=(V_U)_{ji}$ and is experimentally distinguishable from the original process in principle.

\subsection{Results}
\label{sec:res}

\begin{figure}[t] 
	\includegraphics[width=\linewidth]{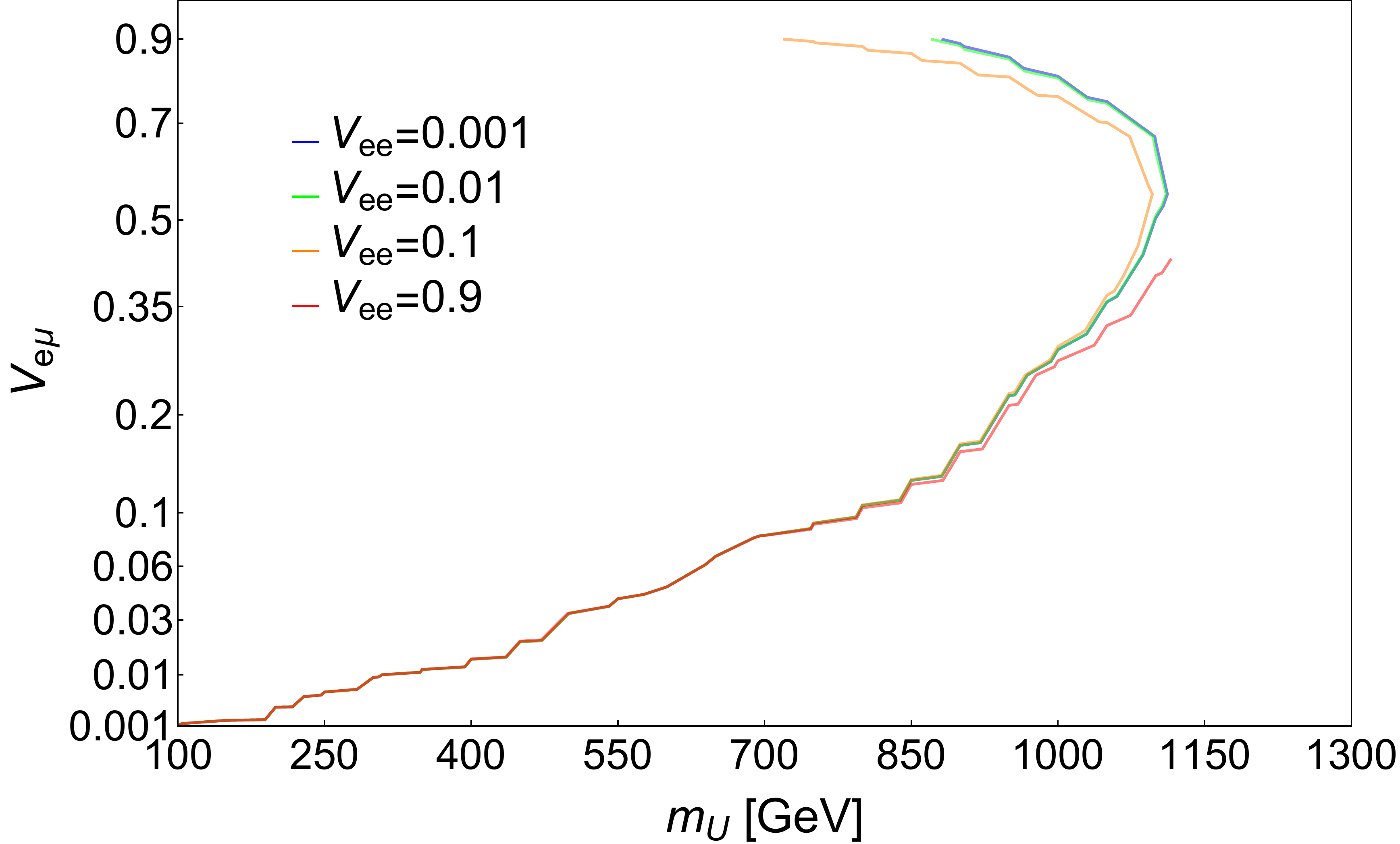}  
	\caption{Occurrence of 3 events for each value of $(V_U)_{ee}$ considered. The region to the left of the curves may be experimentally eliminated with a 95\% confidence level.}
	\label{fig4}
\end{figure}

\begin{figure} 
	\includegraphics[width=1.05\linewidth]{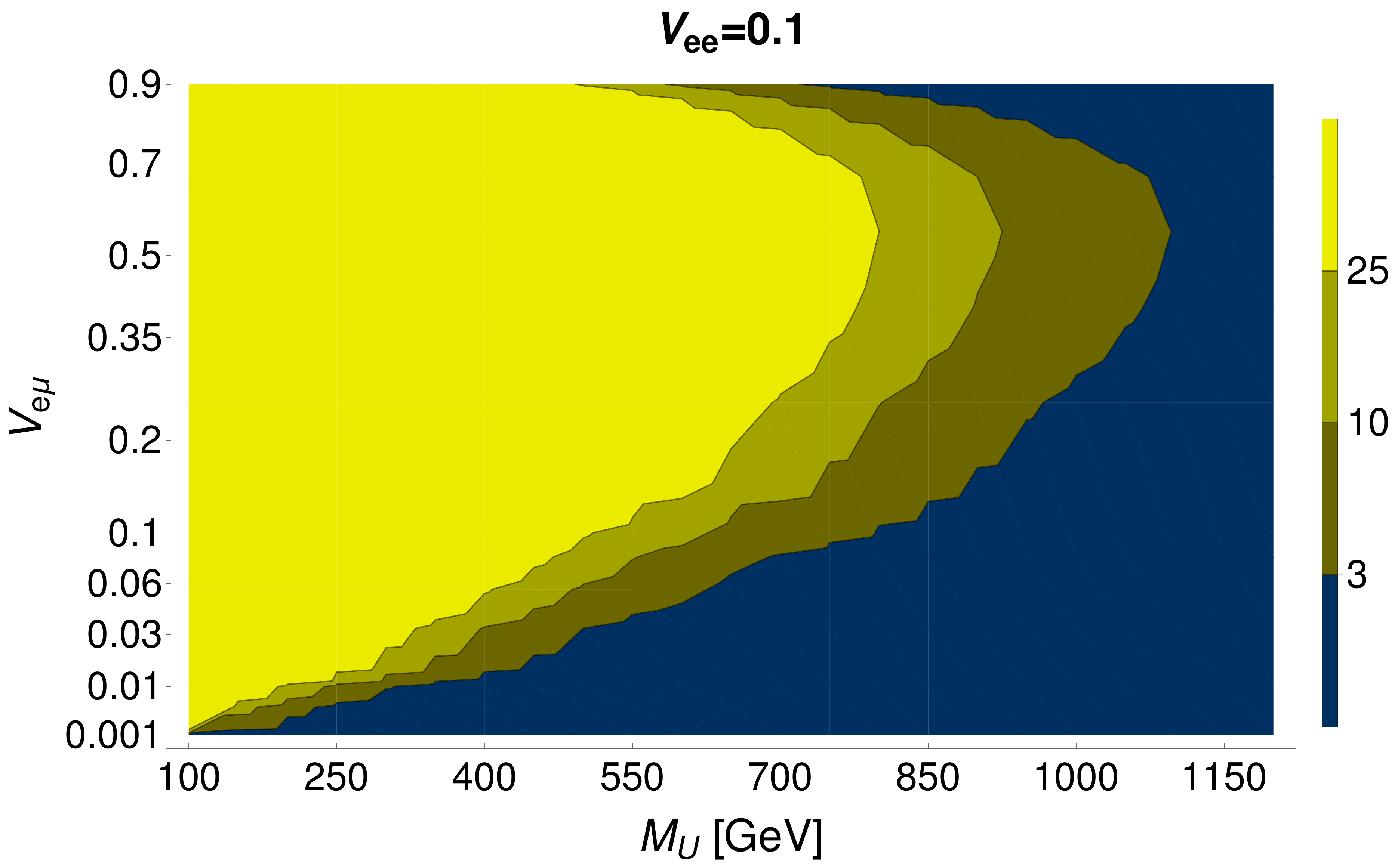}  
	\caption{More detailed example of the behaviour of the number of events. The shown plot is for $(V_U)_{ee}=0.1$.}
	\label{fig5}
\end{figure}

The results are presented in Figs. \ref{fig4} and \ref{fig5}. The $y$-axis is rescaled by a square-root function for better readability of the smaller values of $(V_U)_{e\mu}$. Since there is no background, we present directly contours of number of events instead of confidence levels. The shown contours in Fig.~\ref{fig4} refer to the occurrence of 3 events, so that the region to the left of the curves may, in the respective case, be eliminated with a confidence level of ~95\% (in a Poisson statistic basis and without systematic uncertainties). 

We observe that the contour is roughly identical for all $(V_U)_{ee}$, which happens because the orthogonality constraints obligate the matrix elements to conspire in such a way that when [for fixed $(V_U)_{e\mu}$] $(V_U)_{\mu \mu}$ increases, making the numerator of the cross section larger, the width decreases (roughly) exactly the right amount to make it stay the same.  We see that the highest mass that may be eliminated in the specified conditions is $\SI{\sim 1100}{GeV}$, for $(V_U)_{e\mu}\sim 0.52$.


\section{Conclusions}
\label{sec:con}

We see that at  $\sqrt{s}=\SI{13}{TeV}$ and $\mathcal{L}=\SI{140}{fb^{-1}}$, the
trimuon end state may be adequate to explore the possibil-
ity of doubly charged vectors with masses up to the TeV
scale in favorable cases. Had we chosen to use the SM $WWZ$ vertex for the $UUZ$ interaction, the
trimuon end state may be adequate to explore the possibil-
ity of doubly charged vectors with masses up to the TeV
scale in favorable cases. Had we chosen to use the SM $U$ is $Z$-phobic. By adding other contributions due to the other
particles of a given model, unless a fine-tuned negative
interference happens, our result that a vector bilepton can
be observed at the LHC should not be affected. 

Of course, there might be processes that could, theo-
retically, impose lower limits that are higher than those
obtained in this work for the mass of the doubly charged
bileptons. One example we have noted is the purely
leptonic $\mu\to e\bar{e}e$ decay, which, by an analytical calculation, we observe to be able to explore masses up to 5 TeV if the vector bilepton contribution to that process is predominant~\cite{Machado:2016jzb}. However, in the mentioned reference a
heavy sextet was assumed, and the only d.o.f. active at low
energies were the three scalar triplets. In contrast, if the
d.o.f. of the scalar sextet are active at low energy, light
(neutral or doubly charged) scalars may induce large
contributions to the $\mu\to e\bar{e}e$ decay, possibly relieving the lower limit for the mass of the boson $U$.

Concerning the trimuon events, we recall that many years ago this kind of process was apparently observed in several experiments using
neutrino-nucleon scattering~\cite{Benvenuti:1977zp,Benvenuti:1977fb,Barish:1977bg}. At that time, it was difficult to accommodate these events in electroweak models with a $SU(2) \otimes U(1)$ symmetry, but not in those with a $SU(3)\times U(1)$ one~\cite{Lee:1977qs,Langacker:1977ae}. However, further experiment do not confirm the existence of these events with neutrino energies larger than 100 GeV~\cite{Holder:1977gp}. If they do occur in nature, perhaps they could be observed at the LHC.

The objective of the present paper was to study the contribution of the vector bilepton alone to a hadronic process with lepton flavor violation, so that we could also make a more skeptical assessment of the unavoidable mixing matrix. Nevertheless, we emphasize that a given closed model that contain such bileptons could accommodate a great number of still free parameters, which makes a truly skeptical phenomenological analysis very complicated, and that eventually, a more detailed study, including more d.o.f. of each said model, cannot be avoided.

We conclude that it is worth continuing to study such processes in a model dependent way to see, among other things, if, as we expect, there is no negative interference that can suppress the contribution of the vector bilepton $U^{\pm \pm}$, and we urge the LHC to search for this sort of resonance in the context of such models.

\section*{ACKNOWLEDGMENTS}
M.B. would like to thank CNPq for the financial support and specially Rodolfo M. Capdevilla for many useful discussions. V.P. would like to thank CNPq for partial support and is also thankful for the support of FAPESP funding Grant No. 2014/19164-6.

\end{document}